\documentclass[twocolumn, letter]{jpsj2}
\usepackage{txfonts}
\title{Scalar order: possible candidate for order parameters in skutterudites }

\author{Annam\'aria \textsc{Kiss}\thanks{E-mail address: amk@cmpt.phys.tohoku.ac.jp} and Yoshio
\textsc{Kuramoto}\thanks{E-mail address: kuramoto@cmpt.phys.tohoku.ac.jp}}

\inst{Department of Physics, Tohoku University, Sendai, 980-8578}

\abst{
Phenomenological Landau analysis shows that the properties of ordered phases in some skutterudites are consistently accounted for by a scalar order parameter which preserves the cubic symmetry, even in the ordered phase.
A universal value is found for the anisotropy ratio of the transition temperature in a magnetic field, homogeneous magnetization, and induced staggered magnetization. 
The difference in magnetic behavior 
between PrFe$_4$P$_{12}$ and PrRu$_4$P$_{12}$ near their phase transitions
is explained within a single framework.
For the low-field phase of PrFe$_4$P$_{12}$, the scalar order with the $\Gamma_{1g}$ symmetry can explain 
(i) the absence of field induced dipoles perpendicular to the magnetic field,
(ii) isotropic magnetic susceptibility in the ordered phase, 
(iii) the field angle dependence of the transition temperature, and
(iv) the splitting pattern of the $^{31}$P nuclear magnetic resonance (NMR) spectra. 
It is proposed how the order parameter in SmRu$_4$P$_{12}$ is identified
by NMR analysis of a single crystal. 
}

\kword{skutterudite, PrFe$_4$P$_{12}$, scalar order, magnetic properties, NMR}

\begin{document}
\maketitle

Rare-earth filled skutterudites have been attracting considerable attention 
from both experimental and theoretical sides because of their intriguing behaviors. 
Among them, Pr-based compound
PrFe$_4$P$_{12}$ shows a phase transition at $T_0=6.5$K, which can be seen as sharp anomaly in the magnetic susceptibility\cite{aoki}.
In the ordered phase staggered dipoles are induced by magnetic
field\cite{hao}, which suggests that the order parameter does not break the time-reversal symmetry. 
PrRu$_{4}$P$_{12}$ has a metal-insulator phase transition at $T_{\rm MI}=65$K, and its crystalline electric field (CEF) states show drastic change below $T_{\rm MI}$\cite{iwasa4}, which seems to be described by antiferro-type order of hexadecapole moments with $\Gamma_{1g}$ symmetry\cite{takimoto}.
SmRu$_{4}$P$_{12}$ has also a metal-insulator phase transition at $T_{\rm MI}=16.5$K. 
The nature of the order parameter in phase II is not clear until know. 
An octupolar order with the $\Gamma_{5u}$ symmetry,  which breaks the time-reversal symmetry, has been proposed for this phase\cite{yoshizawa}.  
In fact, nonzero internal field was observed 
below $T_{\rm MI}$ by recent $\mu$SR experiment\cite{hachitani}.

In this paper we propose that both 
PrFe$_4$P$_{12}$ and PrRu$_4$P$_{12}$ have
a scalar-type order parameter with the $\Gamma_{1g}$ symmetry.
We show by phenomenological analysis that the scalar order model explains 
the main properties of PrFe$_{4}$P$_{12}$ including the NMR results: 
(i) the absence of field induced dipoles perpendicular to the magnetic field,
(ii) isotropic magnetic susceptibility in the ordered phase, 
(iii) the field angle dependence of the transition temperature, and
(iv) the splitting pattern of the $^{31}$P NMR spectra. 
As a first step, we concentrate on behaviors at low magnetic fields.
For SmRu$_{4}$P$_{12}$,  
in view of its nearly isotropic behavior in the ordered phase, we propose
another candidate of the octupole order of the $T_{xyz}$ type, which
transforms as a pseudo-scalar with the $\Gamma_{1u}$ symmetry.
It is proposed how the order parameter in SmRu$_4$P$_{12}$ is identified
by NMR for the single crystal. 

Up to the present, the order parameter in PrFe$_4$P$_{12}$ has widely been considered as an antiferro-quadrupolar (AFQ) order of
$\Gamma_3$ moments.
However, this AFQ model fails to account for the
isotropic susceptibility in the ordered phase, for example.
Furthermore, with static $\Gamma_3$ quadrupoles it is difficult to explain
why the field induced staggered dipoles are always parallel to the
field direction, as indicated by neutron diffraction\cite{hao} and NMR \cite{kikuchi1,kikuchi2}. 

A scalar-type order can be of two different kinds: one ($\Gamma_{1u}$) breaks, while the other ($\Gamma_{1g}$) does not break the time-reversal symmetry.
On the other hand, both of them preserve the cubic symmetry even in the ordered phase. 
Therefore, the Landau-type expansion of the free energy ${\cal F}$ contains cubic invariants composed by the magnetic field components. Around a second order phase transition 
%$T\approx T_c$ we can write 
we expand ${\cal F}$ up to fourth order in the order parameter and 
magnetic field as
\begin{eqnarray}
{\cal F}(\psi_{\bf Q},H) =
{\cal F}_0(H)+  \frac 12 a_s [T-T_c(H)]\psi_{\bf Q}^2+
\frac 14 b_s \psi_{\bf Q}^4, 
%\\&& -\frac{1}{2}\chi  H^2,
\label{eq:free3}
\end{eqnarray}
where $\psi_{\bf Q}$ is the staggered component of the scalar order
parameter with ordering vector ${\bf Q}=[1,0,0]$, 
and 
${\bf H}=H(h_x,h_y,h_z)$ is the external magnetic field.
The transition temperature in magnetic field has a dependence:
\begin{align}
T_c(H) = T_0 +\frac 12 t_2 H^2 +\frac 14(t_{4} +  t_{4a} h_{4}) H^4,
\label{Tc}
\end{align}
where $t_2, t_4, t_{4a}$ are expansion coefficients, and 
$h_{4}=h_x^4+h_y^4+h_z^4-3/5$. 
This invariant 
%.
$h_4$
is common in the case of cubic ($O_h$) and the tetrahedral ($T_h$) point group symmetries.
Therefore, our present treatment is valid for both cases. 
%In the last term of expression (\ref{eq:free3}), 
%we assume a ferromagnetic interaction $I$ between the dipoles. 
%Then $\chi$ takes the Curie-Weiss susceptibility in the disordered phase.  We assume the form 
%$1/\chi = 1/\chi_p -I$
%where ${\chi_p} \ (\sim 1/T)$  gives Curie form ${\chi_p}\sim 1/T$ when $I=0$.
The first part ${\cal F}_0(H)$ has a field dependence similar to eq.(\ref{Tc}).

\begin{figure}
\centering
\includegraphics[totalheight=6cm,angle=0]{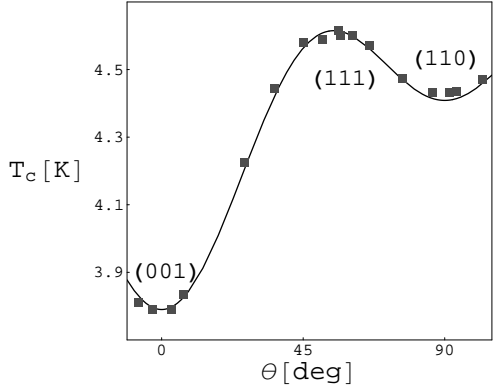}
\caption{$\theta$-dependence of the transition temperature
with $\phi=\pi/4$ choosing coefficients $t_2, t_4$ and $t_{4a}$ to fit $T_c$ %measured at $\theta=0$ and $54.73$. 
for fields along (001) and (111).
Boxes represent the measured result at $H=2.7$T\cite{sakakibara}.}\label{fig:1}
\end{figure}

The anisotropy in eq.(\ref{Tc}) is independent of the microscopic details of the scalar order.  
Using 
$h_4[100]=2/5, 
h_4[110]=-1/10, 
h_4[111]=-4/15$ 
we obtain the ratio
\begin{eqnarray}
\frac{T_c [001]-T_c [111]}
{T_c [110]-T_c [111]} =4.
\label{eq:ratio}
\end{eqnarray}
This relation should hold as long as the magnetic field is weak enough when the scalar order emerges.
Figure~\ref{fig:1} shows 
the transition temperature in PrFe$_{4}$P$_{12}$ measured as a function of field angle\cite{sakakibara} which is defined by
$(h_x,h_y,h_z) =
%\equiv {\bf h}=
( \cos\phi \sin\theta ,\sin\phi \sin\theta ,\cos\theta ) $. 
The anisotropy of eq.(\ref{Tc}) 
with $t_{4a}<0$ provides excellent fit to the observed $T_c$.
This result strongly suggests the scalar order in this compound.

Equation (\ref{eq:free3}) also shows that the magnetization along three principal axes should have the anisotropy ratio:
\begin{align}
(M_{100}-M_{111})/(M_{110}-M_{111})=4,
\label{anisotropic-M}
\end{align}
or, equivalently, $(M_{111}-M_{100})/(M_{110}-M_{100})=4/3$.
The ratio given by eq.(\ref{anisotropic-M}) holds
both in the paramagnetic and ordered phase as long as the magnetic field is weak enough.
Because of the coupling with $\psi_{\bf Q}$,  however, 
the weight of anisotropic part should change at the transition temperature.  It is possible that anisotropy is reversed by keeping the ratio given by eq.(\ref{anisotropic-M}).
Experimentally, the reversed anisotropy is indeed observed at $T=0.3$K \cite{sakakibara}.

We now discuss anisotropy of the staggered magnetization $m_{\bf Q}$ with the order parameter $\Gamma_{1g}$.
The change of the free energy due to $m_{\bf Q}$ is given by
\begin{align}
&{\cal F}(\psi_{\bf Q},m_{\bf Q},H)-{\cal F}(\psi_{\bf Q},H) \nonumber\\
&=\frac 12 a_m m_{\bf Q}^2  +
\psi_{\bf Q}m_{\bf -Q} \left[  
c_{1}H +(c_3 +c_{3a}h_{4})H^3\right],
\label{eq:free1}
\end{align}
where ${\cal F}(\psi_{\bf Q},H)$ is given by eq.(\ref{eq:free3}).
In the presence of $\psi_{\bf Q}$ with symmetry $\Gamma_{1g}$,
$m_{\bf Q}$ is induced by external magnetic field.
Terms with coefficient $c_i$ in expression (\ref{eq:free1}) come from the invariant $\Gamma_{1g}({\bf Q})\otimes \Gamma_{4u}({\bf -Q})\otimes \Gamma_{4u}({\bf 0})$, 
which requires that the induced dipoles are always parallel 
%.
or anti-parallel to the field direction.
%We note that this invariant is absent in the case of order parameter with symmetry $\Gamma_{1u}$, which means the absence of induced staggered dipoles in the ordered phase.
From the condition
$\partial {\cal F}/\partial m_{\bf Q}=0$ we obtain
\begin{eqnarray}
m_{\bf Q}=-\frac{1}{a_m}\left[  
c_1 H+(c_3 +c_{3a }h_{4})H^3 \right]
\psi_{\bf Q},\label{eq:indm}
\end{eqnarray}
which shows explicitly that $m_{\bf Q}$ develops in magnetic field when the order parameter $\psi_{\bf Q}$ is non-zero. 
%.We note that 
In the linear order of magnetic field, there is no anisotropy in the
induced staggered dipoles.  In the third order, the anisotropy ratio is again given by
\begin{align}
(m_{\bf Q}[100]-m_{\bf Q}[111])/(m_{\bf Q}[110]-m_{\bf Q}[111])=4.
\label{anisotropic-m_Q}
\end{align}
At present,  there is no experimental information for the above ratio, since neutron scattering has been done only for $m_{\bf Q}[100]$ and $m_{\bf Q}[110]$ \cite{hao}.
%.
The same anistropy ratio given by eqs.(\ref{eq:ratio}), (\ref{anisotropic-M}) and
(\ref{anisotropic-m_Q}) comes from the property of $h_4$ alone, and is a universal feature for the scalar order.

Let us explain how the difference between PrFe$_4$P$_{12}$ and PrRu$_4$P$_{12}$ can be interpreted in a phenomenological framework. 
The most obvious difference appears in the magnetic susceptibility;
PrFe$_4$P$_{12}$ shows a sharp peak at the transition \cite{aoki}, while
no conspicuous anomaly is seen in PrRu$_4$P$_{12}$. 
In considering the magnetic susceptibility around the zero-field phase transition temperature $T_0$, it is more convenient to use
the Gibbs potential which is obtained from the free energy ${\cal F}$ 
by Legendre transformation.
Namely we obtain
using the homogenous dipole moment ${\bf M}={\bf m}_{\bf 0}$,
\begin{align}
&{\cal G}(\psi_{\bf Q},M)={\cal F}(\psi_{\bf Q},H)+{\bf M}\cdot {\bf H} \nonumber\\
&= {\cal F}(\psi_{\bf Q},H=0)+
\frac 12 a_f (T-T_F)M^2+
\frac 12 \lambda \psi_{\bf Q}^2 M^2,
\end{align}
where ${\cal F}(\psi_{\bf Q},H=0)$ is given by eq.(\ref{eq:free3}) with $H=0$, and 
the last term 
%with $\lambda = -a_s t_2 \chi^2$
represents the coupling between the scalar order and the magnetization.
The term with $a_f$ becomes important if the ferromagnetic instability is close to 
the scalar phase transition: $T_F \lesssim T_0$, as in the case of
PrFe$_4$P$_{12}$.   
Indeed, a positive Curie-Weiss temperature $T_F\approx 3.5$K has been found in experiments\cite{aoki}.
To the contrary, PrRu$_4$P$_{12}$ seems far from the ferromagnetic instability.  In this case, $M$ does not have a significant amplitude, and the coupling term with $\lambda$ is less significant.

%
%The general form of ${\cal G}$ can be written as
%\begin{eqnarray}
%&& \hspace*{-0.6cm}
%{\cal G}_0+\frac{1}{2}A(T-T_0)\psi_{\bf Q}^2+\frac{1}{4}B\psi_{\bf Q}^4\nonumber\\
%&&+\frac{1}{2}C\psi_{\bf Q}^2M^2+\frac{1}{2}LM^2+...,
%\end{eqnarray} 
%We use equation $M=-\partial {\cal F}/\partial H$ to find $H$ as a series in powers of $M$, and from comparison with equation $H=\partial {\cal G}/\partial M$ we find relation between the coefficients appearing in ${\cal F}$ and ${\cal G}$ as $A=a$, $B=b$, $C=c/l^2$ and $L=-1/l=(1/\chi_{p}-K)\equiv \alpha (T-T_F)$.

\begin{figure}
\centering
\includegraphics[width= 0.35\textwidth,angle=270]{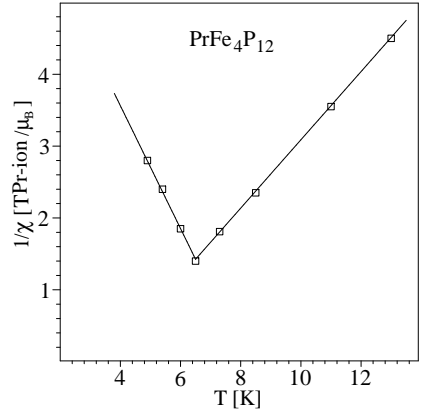}
\caption{Curie-Weiss fit of the inverse susceptibility of PrFe$_4$P$_{12}$.
The parameters are chosen so that 
$\lambda a_s /(a_f b_s)=2.8$. Boxes represent the measured result\cite{aoki}.
}\label{fig:CW-fit}
\end{figure}

Around the transition temperature $T_0$, the magnetic susceptibility 
$\chi_+$ for $T>T_0$ and
$\chi_-$ for $T<T_0$
can be expressed as
\begin{align}
& \chi_{+}^{-1} = a_f (T-T_F), \\
& \chi_-^{-1} =  (a_f -\lambda a_s /b_s)T- a_f T_F+\lambda a_s T_0/b_s,
\label{chi_-}
\end{align}
where both $\chi_{\pm}$ are isotropic and follow the Curie-Weiss law. 
If the coupling with magnetic moment is strong enough, 
we obtain
$
\lambda a_s /b_s > a_f
$,
which leads to a peak in the susceptibility at $T_0$.
Figure \ref{fig:CW-fit} shows comparison between theory and experiment for PrFe$_4$P$_{12}$.  The agreement is excellent with the choice of 
$\lambda a_s /(a_f b_s)=2.8$. 
In consistency with the sharp peak in the susceptibility, 
the phase boundary is suppressed appreciably by magnetic field 
in PrFe$_4$P$_{12}$.

In the opposite limit of negligible $\lambda$, $\chi_-^{-1}$ given by
eq.(\ref{chi_-}) reduces to $\chi_+^{-1}$.  
Hence there is no change at the scalar phase transition.  This limit seems to explain the situation in PrRu$_4$P$_{12}$.
Accordingly, the transition temperature is hardly affected by magnetic field\cite{sekine}. 

%The phase boundary near $T_0$ can be expressed as 
%\begin{eqnarray}
%T_c=T_0-c/aH^2\equiv T_0-a_H H^2.
%\end{eqnarray}
%%In the case of PrRu$_4$P$_{12}$ the phase boundary is not affected so much by increasing magnetic field, which means that $a_H\approx 0$ in this compound giving $c|_{T=T_0} \approx 0$.
%In this case, from expression (\ref{eq:susc2}) we obtain that there is no visible anomaly in the magnetic susceptibility at $T_0$, which is consistent with the experimental situation\cite{sekine}. However, in the case of PrFe$_4$P$_{12}$ the phase boundary is suppressed appreciably by magnetic field, which gives that coefficient $a_H$ and also $c|_{T=T_0}$ is large and positive, which is consistent with the sharp peak in the magnetic susceptibility at $T_0$.

We proceed to analyze $^{31}$P NMR spectra in skutterudites in terms of the scalar order.  We discuss mainly the case of PrFe$_4$P$_{12}$, but also touch on SmRu$_4$P$_{12}$ where the pseudo-scalar $\Gamma_{1u}$
is a candidate of the order parameter.
Around each Pr ion at position (0,0,0), 
there are six P positions ${\bf r}_{1(2)}=(0,v, \pm u)$,
${\bf r}_{3(4)}=(\pm u,0,v)$ and ${\bf r}_{5(6)}=(v, \pm u,0)$, which are 
crystallographically equivalent.   With finite magnetic field, these positions are no longer equivalent, and splitting of NMR lines takes place.
The splitting is of purely magnetic origin, since $^{31}$P ions have no quadrupolar moment with $I=1/2$. 
Therefore, interactions are only between the P nuclear spin $\bf I$ and the dipole $\bf J$ and octupole ${\bf T^\beta, T}_{xyz}$ moments of a Pr ion.

Taking a representative pair at ${\bf r}_{3}$ and ${\bf r}_{4}$, we write down
the hyperfine interaction characterized by energies 
$e_{k,l}^{(d)}$ for the dipoles and $e_{k,l}^{(o)}$ for octupoles with $k,l$ distinguishing independent components\cite{sakai}.
The invariant form of the interaction contains terms such as
$I_x J_z$ and $I_y T_{xyz}$ since there is only a mirror symmetry against the $xz$-plane for the pair.
The hyperfine interaction is given by
\begin{eqnarray}
&& H_{\rm hf}(3,4) = I_x[e_{1,1}^{(d)} J_x\pm
e_{1,2}^{(d)}J_z+e_{1,1}^{(o)}{T}^{\beta}_x\pm e_{1,2}^{(o)}{
T}^{\beta}_z]\nonumber\\
&&+I_y[e_{2,1}^{(d)}J_y+e_{2,1}^{(o)}{T}^{\beta}_y\pm e_{2,2}^{(o)}T_{xyz}]\nonumber\\
&&+I_z[e_{3,1}^{(d)}J_z\pm e_{3,2}^{(d)}J_x+e_{3,1}^{(o)}{T}^{\beta}_z\pm
e_{3,2}^{(o)}{T}^{\beta}_x]\,,\label{eq:int1}
\end{eqnarray}
where the negative sign corresponds to P ion at position ${\bf r}_{4}$.
The interaction for other pairs can be obtained from eq.(\ref{eq:int1}) by using proper rotational operations.

\begin{figure}
\centering
\includegraphics[width= 0.46\textwidth]{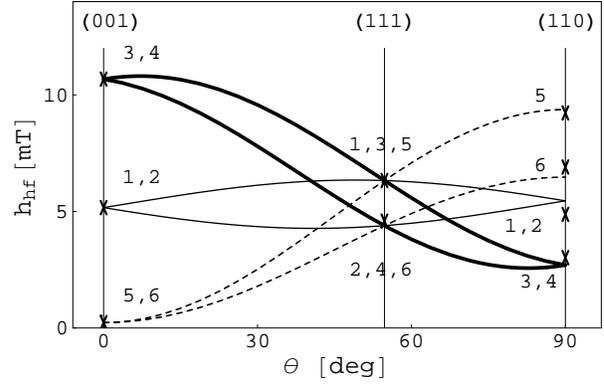}
\caption{Theoretical fitting (six lines) to experimental $^{31}$P NMR results\cite{kikuchi1} (crosses) in PrFe$_4$P$_{12}$ at magnetic field $H=4$T and temperature  $T=50$K as a function of field direction with $\phi=\pi/4$.  
The numbers indicate the P positions.}\label{fig:2}
\end{figure}

Let us discuss first the case of PrFe$_4$P$_{12}$.
In $T_h$ symmetry, the dipole moment ${\bf J}$ corresponds to $\Gamma_{4u}^{(1)}$ and the octupole moment ${\bf T}^{\beta}$ to $\Gamma_{4u}^{(2)}$, 
but they are mixed due to the lower symmetry. Therefore, both are induced by the external magnetic field.
In the disordered phase ($\psi_{\bf Q}=0$),
the homogenous moments are induced as 
${\bf m}_{\bf 0} (= {\bf J}_{\bf 0}) =M{\bf H}/|{\bf H}|$ and ${\bf T}^{\beta}_{\bf 0}=T{\bf H}/|{\bf H}|$, where $M$ and $T$ are 
%.expectation values 
magnitudes
at a given temperature and magnetic field.
They cause the splitting of the P NMR line depending on the field direction.
Since the external magnetic field is much larger than the hyperfine
field,  we assume
$(I_x,I_y,I_z)\propto (h_x,h_y,h_z)$. 
%For field direction $(001)$, for example,  
%the line splittings are described in units of energy by
%\begin{eqnarray}
%h_{\rm hf}({\bf r}_{1(2)}) &=& \left(K_{0,d} e_{1,1}^{(d)}+K_{0,o} e_{1,1}^{(o)}\right)H, \nonumber\\
%h_{\rm hf}({\bf r}_{3(4)}) &=& \left(K_{0,d} e_{3,1}^{(d)}+K_{0,o} e_{3,1}^{(o)}\right)H, \nonumber\\
%h_{\rm hf}({\bf r}_{5(6)}) &=& \left(K_{0,d} e_{2,1}^{(d)}+K_{0,o} e_{2,1}^{(o)}\right)H.\label{eq:paral}
%\end{eqnarray}
In the case of UFe$_4$P$_{12}$, classical approximation of the dipolar field
shows good qualitative agreement with the measured NMR spectra\cite{tokunaga}.
However in PrFe$_4$P$_{12}$,  we checked that 
the approximation leads to a large deviation from 
the measured result\cite{kikuchi1}.
%.
Hence we fix the parameters in eq.(\ref{eq:int1}) in a phenomenological manner.
%to display the angle dependence explicitly.  
We define the hyperfine field $h_{\rm hf}$ so that 
$H_{\rm hf} = \gamma_n I h_{\rm hf}=I(f-f_0)$, where $\gamma_n$ is the nuclear gyromagnetic ratio of $^{31}$P, $f$ is the resonance frequency and $f_0$ is the zero shift. 
We write $h_{\rm hf} (1, 2)\equiv g_1$, $h_{\rm hf} (3,4)\equiv g_2$ and $h_{\rm hf} (5,6)\equiv g_3$ for ${\bf H}\parallel (001)$, and $h_{\rm hf} (1,3,5)\equiv k_1$ for ${\bf H}\parallel (111)$.
%., and then
We obtain
\begin{align}
& h_{\rm hf} (1, 2) = g_3h_x^2 + g_2h_y^2 + g_1h_z^2 \pm g_4 h_yh_z, \nonumber\\
& h_{\rm hf} (3, 4) = g_1h_x^2 + g_3h_y^2 + g_2h_z^2 \pm g_4 h_zh_x,  \nonumber\\
& h_{\rm hf} (5, 6) = g_2h_x^2 + g_1h_y^2 + g_3h_z^2 \pm g_4 h_xh_y, 
\end{align}
where $g_4=3k_1-g_1-g_2-g_3$ and parameters $g_i \ (1\le i\le 4)$ are linear combination of $e_{k,l}^{(\alpha)} \ (\alpha=d,o) $ times $M$ or $T$.
We determine the four parameters $g_i $ by fitting to the observed three lines for ${\bf H}\parallel (001)$ and the three degenerate ones from P1, P3, P5 for ${\bf H}\parallel (111)$.  
%.Then the computed spectrum for ${\bf H}\parallel (110)$ is found to be in reasonable agreement with experimental results.
Figure \ref{fig:2} shows our results for the fitting in the disordered phase. 
The values for parameters 
$g_i$ are summarized in Table~\ref{tab:1}.
The result for ${\bf H}\parallel (110)$ is a consequence of the form of the hyperfine interaction given by eq.(\ref{eq:int1}), which is independent of the microscopic details of the model. 
%.
The spectrum computed for ${\bf H}\parallel (110)$ is found to be in reasonable agreement with experimental results.   Since eq.(\ref{eq:int1}) includes only nearest neighbor interaction between the Pr and P ions, 
the slight deviation between theory and experiment is ascribed to effects of distant Pr-P pairs.

\begin{figure}[h]
\centering
\includegraphics[width= 0.34\textwidth]{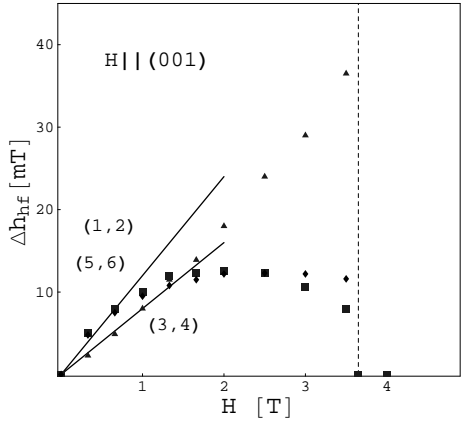}
\caption{Magnetic field dependence in linear order in $H$ (solid line) of the extra line splittings for field direction (001) in PrFe$_4$P$_{12}$. Symbols are the measured result at $T=2$K\cite{kikuchi2} and numbers indicates the P positions. The parameters used are shown in Table~\ref{tab:1}.}
\label{fig:3}
\end{figure}

Let us now consider the case of $\psi_{\bf Q}\neq 0$ with the symmetry $\Gamma_{1g}$.
%In $T_h$ symmetry, the dipole moment ${\bf J}$ corresponds to $\Gamma_{4u}^{(1)}$ and the octupole moment ${\bf T}^{\beta}$ to $\Gamma_{4u}^{(2)}$, but they are mixed due to the lower symmetry.
We restrict our discussion to the case of low magnetic fields, and expand the quantities in linear order of magnetic field.
Thus we write the staggered dipoles and octupoles induced by the magnetic field as
${\bf m}_{\bf Q} = K_{1,d}\psi_{\bf Q}{\bf H}$ and 
${\bf T}^{\beta}_{\bf Q} =K_{1,o}\psi_{\bf Q}{\bf H}$.
As a result,
Pr positions $(0,0,0)$ and $(1/2,1/2,1/2)$ become inequivalent, and
extra splitting of the NMR lines develops below $T_c$.
In magnetic field along (001), for example,  the extra splittings of the three main lines are described by
\begin{align}
&\Delta h_{\rm hf}(1,2) =  \psi_{\bf Q} \left(K_{1,d} e_{1,1}^{(d)}+K_{1,o} e_{1,1}^{(o)}\right)H \equiv a_1 H,\nonumber\\
&\Delta h_{\rm hf}(3,4) =  \psi_{\bf Q} \left(K_{1,d} e_{3,1}^{(d)}+K_{1,o} e_{3,1}^{(o)}\right)H\equiv a_2 H,\nonumber\\
&\Delta h_{\rm hf}(5,6) =  \psi_{\bf Q} \left(K_{1,d} e_{2,1}^{(d)}+K_{1,o} e_{2,1}^{(o)}\right)H\equiv a_3 H.\nonumber
%\label{eq:paral2}
\end{align}
We determine magnitudes of parameters
$a_i \ (i=1,2,3$)
 so as to reproduce the corresponding experimental results\cite{note1}.
%.
Figure \ref{fig:3} shows the result of fitting together with experimental results.
It is obvious that the experimental results deviate from the linear behavior for $H\gtrsim 1$T.  
The different field dependence for the splittings $\Delta h_{\rm hf}(1,2)(\approx \Delta h_{\rm hf}(5,6))$ and $\Delta h_{\rm hf}(3,4)$ is intriguing.  It seems hard to understand the difference
without consideration of induced staggered octupoles.   
The analysis with non-linear effects of magnetic field
is rather complicated, and will be presented in a separate publication.

\begin{table}
\caption{Choice for the parameters to describe the measured NMR results.} 
\label{tab:1}

\medskip\centering
\begin{tabular}{|c|c|c|c|}
\hline $g_1$ & $g_2$ & $g_3$ & $k_1$ \\ 
\hline $5.2$\hspace*{0.4mm}mT  & $10.7$\hspace*{0.4mm}mT & $0.2$\hspace*{0.4mm}mT & $6.3$\hspace*{0.4mm}mT  \\ \hline\hline
 $a_1=a_3$ & $a_2$ & $c_1$ & $ c_2$  \\ 
\hline  $1.2 \times 10^{-2}$ & $0.8 \times 10^{-2}$ & $1.14 \times 10^{-2}$ & $0.99 \times 10^{-2}$   \\ \hline
\hline    $d_1$ & $d_2$ & $d_3$ & $d_4$   \\ 
\hline    $1.0 \times  10^{-2}$ & $1.2 \times 10^{-2}$ & $1.1 \times 10^{-2}$ & $0.9 \times 10^{-2}$  
\\ \hline
\end{tabular}
\end{table}

For ${\bf H}\parallel (111)$, we define the parameters 
$c_1$ for $\Delta h_{\rm hf}(1,3,5)$ and 
$c_2$ for $\Delta h_{\rm hf}(2,4,6)$ in a way 
%.
analogous to $a_i$.
The simple splitting of lines $(1,3,5)$ and $(2,4,6)$ in the ordered phase can be explained with scalar order, but not with quadrupolar order.
The splittings $\Delta h_{\rm hf}(1,3,5)$ and $\Delta h_{\rm hf}(2,4,6)$ have almost identical field dependence with a tiny deviation at low fields\cite{kikuchi1}.
We fix the parameter $c_1$ to reproduce the experimental result, and then parameter $c_2$ is determined as $c_2=(2a_1+2a_2+2a_3-3c_1)/3$. For ${\bf H}\parallel (110)$, we define
$d_1$ for $\Delta h_{\rm hf}(1,2)$,
$d_2$ for $\Delta h_{\rm hf}(3,4)$,
$d_3$ for $\Delta h_{\rm hf}(5)$, and
$d_4$ for $\Delta h_{\rm hf}(6)$.
It turns out that all $d_i$ can be fixed by the parameters $a_j$ and $c_1$ as
$d_1 =(a_2 + a_3)/2, \ 
d_2 =(a_1 + a_3)/2, \ 
d_3 =(3c_1-a_3)/2$ and $d_4=(2a_1+2a_2+a_3-3c_1)/2$. 
Namely, experimental results along (110) should
be reproduced without further adjustable parameters provided eq.(\ref{eq:int1}) applies to the actual system.
%., which is indeed the case.
Experimental values with superscript $e$ are 
$
c_2^{e}=1.0\times 10^{-2}, \
d_1^{e}\sim d_3^{e} = 0.85\times 10^{-2}, \
d_2^{e}=1.2 \times 10^{-2}, \
d_4^{e}=1.1 \times 10^{-2}$ [\citen{kikuchi1}],
which show reasonable agreement with corresponding theoretical values shown 
in Table~\ref{tab:1}.
%.%
We expect qualitatively similar pattern for the NMR spectra in the ordered phase of PrRu$_4$P$_{12}$,  since its order parameter should also be a scalar.

We now discuss briefly the 
expected splitting pattern of the NMR spectra in the case of $T_{xyz}$ octupolar order.
When the magnetic field is applied along $(001)$, the hyperfine interaction for different pairs is given as
\begin{align}
& H_{\rm hf}(1,2) = e_{1,1}^{(d)} I_z J_z, \nonumber\\
& H_{\rm hf}(3,4) = e_{3,1}^{(d)} I_z J_z, \nonumber\\
& H_{\rm hf}(5,6) = e_{2,1}^{(d)} I_z J_z\pm e_{2,2}^{(o)}I_z T_{xyz}.
\label{eq:paral4}
\end{align}
%Thus 
The $T_{xyz}$ octupolar order causes the splitting of lines $(5,6)$.
With the octupole order parameter $\psi_{\bf Q}$ we obtain $\Delta h_{\rm hf}(5)=  \Delta h_{\rm hf}(6)\propto e_{2,2}^{(o)}\psi_{\bf Q}$.
The splitting occurs also for the octupole  
ordering vector ${\bf q}=0$ because $h_{\rm hf}(5)-h_{\rm hf}(6)\propto e_{2,2}^{(o)}\psi_{\bf 0}\ne 0$ in this case.
Then four NMR lines appear in the ordered phases with ${\bf q}={\bf Q}$ as well as ${\bf q}=0$. 
However, in the case of field along $(111)$, 
doubling of each line occurs with ${\bf q}={\bf Q}$,
while no extra splitting 
is expected with ordering vector ${\bf q}=0$.
The former splitting pattern is similar to the case of PrFe$_4$P$_{12}$.

In this paper we have considered the characteristic features of scalar orders with symmetries $\Gamma_{1g}$ and $\Gamma_{1u}$ in weak magnetic fields.
We have found the universal anisotropy ratio in weak magnetic field by phenomenological Landau-type analysis.
We conclude that the scalar order scenario with symmetry $\Gamma_{1g}$ can explain the known properties of PrFe$_4$P$_{12}$ consistently.
We have shown that the splitting pattern of the $^{31}$P NMR spectra in the disordered and ordered phases can also be described within this framework. 
%The field induced octupoles beside the dipoles seem to be indispensable to %account for the characteristic features of the NMR spectra. 
We 
%.derived also 
have predicted the $^{31}$P NMR spectra in the case of $T_{xyz}$ octupole order for fields along $(001)$ and $(111)$.
This octupole moment transforms as a pseudo-scalar with the $\Gamma_{1u}$ symmetry and 
can be a good candidate for the order parameter in SmRu$_4$P$_{12}$ below the metal-insulator transition.
%Further experiments like NMR on single crystal or the measurement of the transition temperature as a function of field angle are important to clarify the nature of the order parameter in this compound.

\section*{Acknowledgment}

The authors are grateful to K. Iwasa, J. Kikuchi, M. Takigawa, D. Kikuchi, T. Tayama, and T. Sakakibara for showing their experimental results prior to publication, and O. Sakai for inspiring discussions.


\begin{thebibliography}{99}
\bibitem{note1}
For the fitting in linear order in $H$ we use the second lowest point of the measured spectra instead of the first one
because of its ambiguity due to the finite linewidth and small value of the magnetic field.
\end{thebibliography}
\end{document}